\documentclass[proof]{WileyASNA-v1}

\defcitealias{GAIADR3}{Gaia Collaboration 2022}

\articletype{Original Article}%

\received{31 July 2024}
\revised{19 December 2024}
\accepted{23 December 2024}

\begin{document}

\title{Where in the Milky Way do exoplanets preferentially form?}

\author[1]{Joana Teixeira*}

\author[1,2]{Vardan Adibekyan}

\author[3,4]{Diego Bossini}

\authormark{Teixeira \textsc{et al}}

\address[1]{Instituto de Astrof\'isica e Ci\^encias do Espa\c{c}o, Universidade do Porto, CAUP, Rua das Estrelas, 4150-762 Porto, Portugal}

\address[2]{Departamento de F\'{\i}sica e Astronomia, Faculdade de Ci\^encias, Universidade do Porto, Rua do Campo Alegre, 4169-007 Porto, Portugal}

\address[3]{Department of Physics and Astronomy G. Galilei, University of Padova, Vicolo dell'Osservatorio 3, I-35122, Padova, Italy}

\address[4]{INAF-Osservatorio Astronomico di Padova, Vicolo dell'Osservatorio 5, I-35122 Padova, Italy}

\corres{*\email{joanaraquel.teixeira00@gmail.com}}


\abstract{Exoplanets are detected around stars of different ages and birthplaces within the Galaxy. The aim of this work is to infer the Galactic birth radii ($r_\text{birth}$) of stars and, consequently, their planets, with the ultimate goal of studying the Galactic aspects of exoplanet formation. We used photometric, spectroscopic, and astrometric data to estimate the stellar ages of two samples of stars hosting planets and, for comparison, a sample of stars without detected planets. The $r_\text{birth}$ of exoplanets were inferred by projecting stars back to their birth positions based on their estimated age and metallicity [Fe/H]. We find that stars hosting planets have higher [Fe/H], are younger, and have smaller $r_\text{birth}$ compared to stars without detected planets. In particular, stars hosting high-mass planets show higher [Fe/H], are younger, and have smaller $r_\text{birth}$ than stars hosting low-mass planets. We show that the formation efficiency of planets, calculated as the relative frequency of planetary systems, decreases with the galactocentric distance, which relationship is stronger for high-mass planets than for low-mass planets. Additionally, we find that (i) the formation efficiency of high-mass planets increases with time and encompasses a larger galactocentric distance over time; (ii) the formation efficiency of low-mass planets shows a slight increase between the ages of 4 and 8 Gyr and also encompasses a larger galactocentric distance over time; and (iii) stars without detected planets appear to form at larger galactocentric distances over time. We conclude that the formation of exoplanets throughout the Galaxy follows the Galactic chemical evolution, for which our results are in agreement with the observed negative interstellar medium (ISM) metallicity gradient and its enrichment and flattening with time at any radius.}


\jnlcitation{\cname{%
\author{Teixeira, J.}, 
\author{Adibekyan, V.}, and
\author{Bossini, D.} (\cyear{2025}), 
\ctitle{Where in the Milky Way Do Exoplanets Preferentially Form?}. \cjournal{Astron. Nachr.} \cvol{e20240076}.}}


\maketitle


\section{Introduction} \label{sec:intro}

Over 5500 exoplanets have been discovered thus far, implying that planet formation is a universal process that occurs around the majority of stars \citep[e.g.,][]{Cassan_2012}. The discovered exoplanets exhibit a large diversity of properties, some of which are intrinsically correlated to those of their host stars. These correlations give us important constraints about planetary formation and evolution.
In particular, observations have shown a positive correlation between stellar metallicity [Fe/H], and the presence of giant planets \citep[e.g.,][]{Gonzalez_1997,Santos_2000,2005ApJ...622.1102F,2007ARA&A..45..397U,Mortier}.
This relation is expected by core accretion models \citep{Pollack1984}. Therefore, stellar metallicity plays a crucial role in the formation and occurrence of exoplanets \citep[e.g.,][]{HeavyMetalRules}.

Stars imprint in their atmospheres the metallicity of the interstellar medium (ISM) at the time and place they are formed.
As such, studies have been exploring how the Galactic chemical evolution impacts the formation and evolution of planetary systems, and how the origin of stars affects the formation of planetary building blocks across the Galaxy \citep[e.g.,][]{Tc_Adibekyan2014,  Adibekyan2015,Tc_Adibekyan2016, GCE_terrestrialexoplanets2020,Swastik2022, Baba2023, Nielsen2023,Cabral2023, Boettner2024}.

The chemical composition of the Galaxy varies through time as a function of the galactocentric radius.
Depending on stellar mass, different elements are returned to the ISM through core-collapse supernovae (SNe) \citep{Matteucci_2021}. In particular, SNe type II is the main contributor of $\alpha$-elements to the ISM on short timescales. On the other hand, SNe type Ia is the major producer of heavy metals into the ISM on longer timescales, and, therefore, younger stars are metal-richer compared to their older counterparts.
Furthermore, today is observed a radial negative metallicity gradient in the Galactic disc of $\sim$ $-0.07$ dex kpc$^{-1}$ \citep[e.g.,][]{Minchev}, which is believed to be the result of an inside-out formation of the Galactic disc \citep{Matteucci_Francois1989}. Consequently, more stars are formed per unit of time in the inner disc, leading to a higher enrichment of the ISM in this region.

The locations at which planetary systems are observed today do not represent the environments from which they were formed. Stars radially migrate from their birthplace around the Galaxy during their lifetime \citep{Sellwood_2002}.
\cite{Minchev} developed a method to infer the Galactic birthplaces of stars, relying only on precise [Fe/H] and age estimations. This method consists of comparing the chemical abundance of a star of a particular age with the chemical composition of the ISM at the time of the star’s formation.
The authors found that the time evolution of the ISM metallicity at the solar radius, [Fe/H]$_\text{ISM}(r_\odot)$, and the time evolution of the ISM metallicity gradient, d[Fe/H]$_\text{ISM}$/d$r$, are best described by logarithmic functions. Additionally, they found that at early times, the variation of the slope was $-0.15$ dex kpc$^{-1}$, flattening towards redshift zero to the present day $-0.07$ dex kpc$^{-1}$.

In this work, we infer the Galactic birth radii ($r_\text{birth}$) of stars and, consequently, their planets, with the ultimate goal of understanding the galactic aspect of exoplanet formation and evolution. The paper is organised as follows: Section~\ref{sec:data} presents the data and samples used in this work. Section~\ref{sec:ages} describes the methods used to infer stellar ages and $r_\text{birth}$. Section~\ref{sec:results} shows the results obtained in this work. Section~\ref{sec:conclusion} provides a discussion of the results and the conclusion.

\section{Samples and Data Collection}					\label{sec:data}

We started by selecting a sample of stars with planets from the SWEET-Cat catalogue \citep{SWEETCat2013,SWEETCat2021} and a comparison sample of stars with and without detected planets from the HARPS-GTO database \citep{HARPS-GTO_2012,HARPS-GTO_2017}. Both catalogues provide stellar parameters derived in a homogeneous way (marked as "SWflag=1" in SWEET-Cat), from which we retrieved the effective temperature ($T_\text{eff}$), trigonometric surface gravity ($\log \textsl{g}$), [Fe/H], and their respective uncertainties. We added quadratically a factor of 0.04 dex in [Fe/H], 60 K in $T_\text{eff}$, and 0.1 dex in $\log \textsl{g}$ to the precision uncertainties of these quantities, following \cite{Sousa2011}.

We collected the apparent magnitudes ($m_\lambda$) from the Two Micron All-Sky Survey \citep[2MASS;][]{2MASS}, where we selected only the $K_s$ magnitudes, and from Gaia Data Release 3 \citepalias[DR3;][]{GAIADR3}, where we selected the $G$ magnitudes, the blue $G_\text{BP}$, and red $G_\text{RP}$ passbands and their respective mean flux over error ($\Delta F/F$). The uncertainties for each Gaia passband were computed by error propagation, adding quadratically their respective zero point uncertainty as provided in \cite{GaiaEDR3}.

We selected only stars with relative uncertainties in parallax, $p$, smaller than 10\% from Gaia DR3, since a significant uncertainty in parallax measurement can lead to a notable bias in the distance estimation \citep{distance_from_parallax}. In this work, we estimated distances simply as the inverse of parallax.

The properties of the exoplanets were extracted from the NASA Exoplanet Archive (NEA)\footnote{\url{https://exoplanetarchive.ipac.caltech.edu/}}, for which we selected only planets with $M_p<11$ M$_J$, which corresponds approximately to the deuterium-burning mass limit for brown dwarfs \citep{BD_limit}. 

The SWEET-Cat sample consists of 670 stars hosting 872 planets located within 2 kpc from the Sun. These stars are mainly FGK-type stars with $T_\text{eff}$ between 4103 and 7201 K and $\log \textsl{g}$ between 2.09 and 4.88 dex, however, most of the stars are in the main sequence evolutionary stage, and only a very few are evolved stars. Regarding [Fe/H], the stars cover a range between -0.69 and 0.55 dex.

\begin{figure}
    \centering
    \includegraphics[scale=0.25]{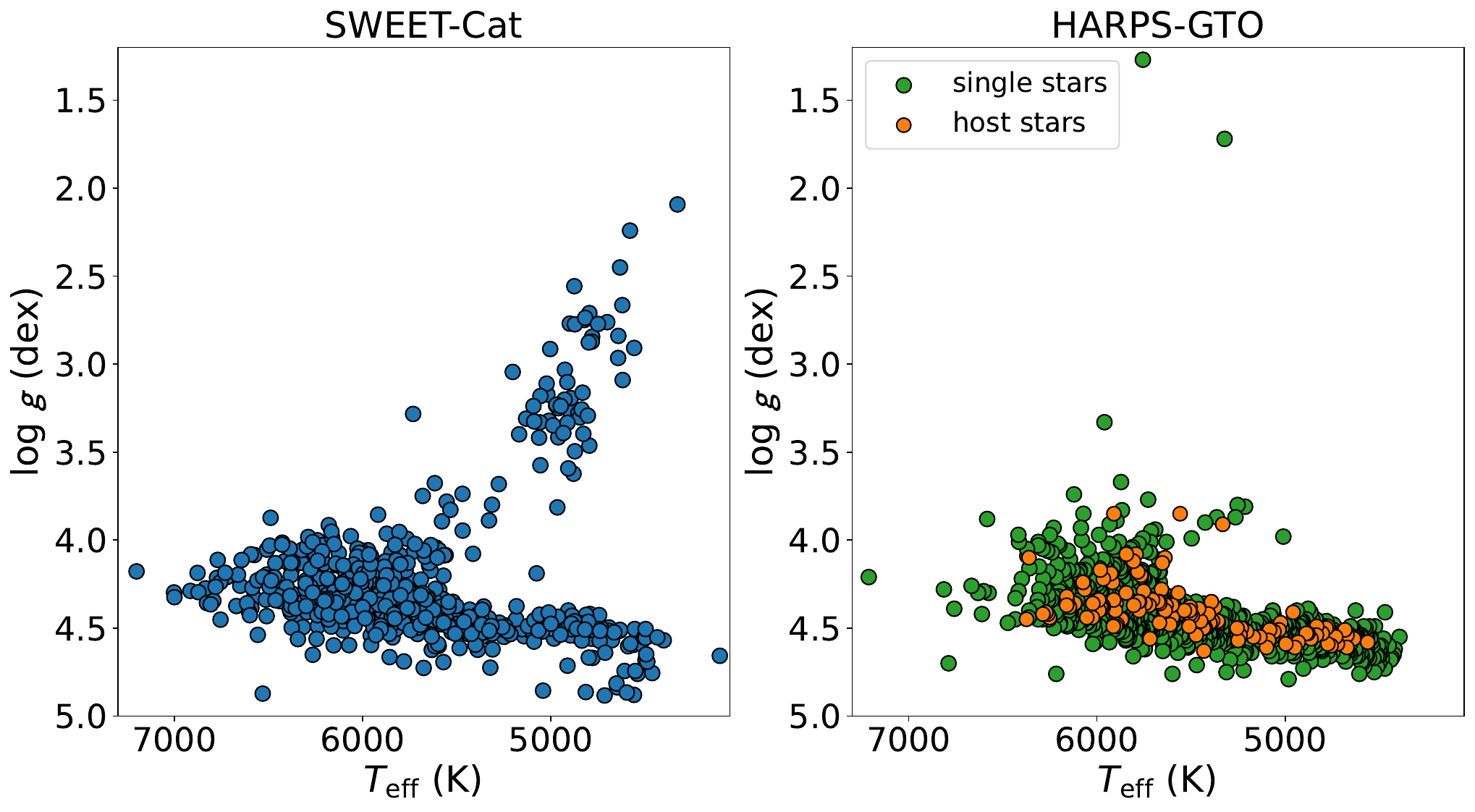}
    \caption{The $\log \textsl{g}$ -- $T_\text{eff}$ diagram for SWEET-Cat host stars (\textit{left panel}) and for HARPS-GTO host and single stars (\textit{right panel}).}
    \label{fig:teff_logg}
\end{figure}

The HARPS-GTO sample is composed of 818 stars without detected planets (hereafter, single stars) and 120 stars hosting 177 planets, lying within 500 pc from the Sun. The stars without planets cover the following range of parameters: $4393 < T_\text{eff} < 7212$ K, $1.27 < \log \textsl{g} < 4.79$ dex, and $-1.39 < \text{[Fe/H]} < 0.45$ dex. The host stars cover the following range of parameters: $4562 < T_\text{eff} 
< 6374$ K, $3.85 < \log \textsl{g} < 4.63$ dex, and $-0.62 < \text{[Fe/H]} < 0.37$ dex. Figure~\ref{fig:teff_logg} shows the SWEET-Cat and HARPS-GTO stars in the $\log \textsl{g}$ -- $T_\text{eff}$ diagram.

\section{Stellar ages and Galactic birth radii}          \label{sec:ages}

We estimated the stellar ages ($\tau$) with PARAM \citep{PARAM2006,PARAM2014,PARAM2017} using a set of isochrones from PARSEC \citep{PARSEC}. As input parameters, we set $T_\text{eff}$, [Fe/H], and luminosity ($L$). As output, PARAM computes a posterior probability density function (PDF) of the model parameters, and then the median, mode, and 68th percentile are calculated from each PDF.

For each star, we derived two different luminosities, $L^G_\text{bol}$ and $L^{K_s}_\text{bol}$, from $G$ and $K_s$ magnitudes, respectively. We chose to use both bands because, while the $K_s$-band is less sensitive to interstellar extinction, and thus, it is expected that the $L_{K_s}$ results are more accurate, our stars emit more radiation in the $G$-band, making the $L_G$ luminosities potentially more precise.
To derive the bolometric stellar luminosity at each wavelength $\lambda$, we used the following expression
\begin{equation}
    \frac{L_\mathrm{bol}^\lambda}{L_\odot}=10^{-0.4\left(m^0_\lambda + A_\lambda - 5 \log_{10}\left(\frac{1/p}{10 \ \text{pc}} \right) + BC_\lambda -M_{\mathrm{bol},\odot}\right)},
\end{equation}
where $m_\lambda^0$ is the intrinsic magnitude, $A_\lambda$ is the interstellar extinction, $BC_\lambda$ is the bolometric correction calculated using the YBC\footnote{\url{https://sec.center/YBC/}} code \citep{YBC}, and $M_{\mathrm{bol},\odot}$ is the absolute bolometric magnitude for the nominal solar luminosity (+4.74 mag) according to \cite{IAU_2015}.
We also corrected the reddening caused by interstellar dust using the STructuring by Inversion the Local Interstellar Medium \citep[STILISM;][]{STILISM2014,STILISM2019}. The uncertainties of the parameters $m_\lambda^0$ and $L_{\mathrm{bol}}^\lambda$ were calculated through bootstrapping.

We inferred two different ages, $\tau_G$ and $\tau_{K_s}$, for each star corresponding to the input luminosities.
To combine the age estimates into one, we had to take into account the age limit used to infer the Galactic birth radii: 13.3 Gyr. As such, stars with [$\tau_G$, $\tau_{K_s}$] > 13.3 Gyr were removed from the samples. In the case where [$\tau_G$, $\tau_{K_s}$] < 13.3 Gyr, we generated random samples using the PDFs of $\tau_G$ and $\tau_{K_s}$ obtained with PARAM. These were then combined into a single array, forming a joint random distribution. In cases where only one estimate fell below the age limit, for consistency, we also generated a random sample from the PDF of the age estimate below the limit. The final age of each star was determined as the median of the distributions, with uncertainties measured as one standard deviation of the mean of the distribution.

Our final sample consists of 548 stars hosting 703 planets from the SWEET-Cat catalogue, and 83 stars hosting 120 planets and 491 single stars from the HARPS-GTO database.

To infer the $r_\text{birth}$ of stars, we adopted the same method as \cite{Minchev}, using [Fe/H] and age estimations. We choose this method because it consists of projecting stars back to their birth position according to only their [Fe/H] and age measurements, with no kinematical data required, which makes this semi-empirical and model-independent approach intuitive and easy to interpret. In practice, the $r_\text{birth}$ was inferred from the [Fe/H]--$r$ relation for a given age. Their uncertainties were determined through bootstrapping.

\section{Results}          \label{sec:results}

We divided the stars into three groups according to the planetary mass they host: stars hosting high-mass planets with $M_p \ge$ 50 M$_\oplus$ (hereafter, HMPHs: 475 stars hosting 544 planets in the SWEET-Cat and 68 stars hosting 81 planets in the HARPS-GTO), stars hosting low-mass planets with $M_p \le$ 30 M$_\oplus$ (hereafter, LMPHs: 80 stars hosting 141 planets in the SWEET-Cat and 19 stars hosting 37 planets in the HARPS-GTO), and stars hosting exclusively low-mass planets with $M_p \le$ 30 M$_\oplus$ (hereafter, ELMPHs: 61 stars hosting 107 planets in the SWEET-Cat and 13 stars hosting 24 planets in the HARPS-GTO).
These lower and upper mass limits for high- and low-mass planets were chosen according to predicted values from observations and core accrection models \citep[e.g.,][]{Mayor2011,Mordasini2009}. Nevertheless, these values essentially correspond to the observed mass gap presented in mass-period and mass-radius diagrams \citep{HeavyMetalRules}.
We note that some planetary systems can appear in both HMPHs and LMPHs groups.

\subsection{Metallicity distributions} \label{feh}

\begin{figure*}
    \centering
    \includegraphics[scale=0.31]{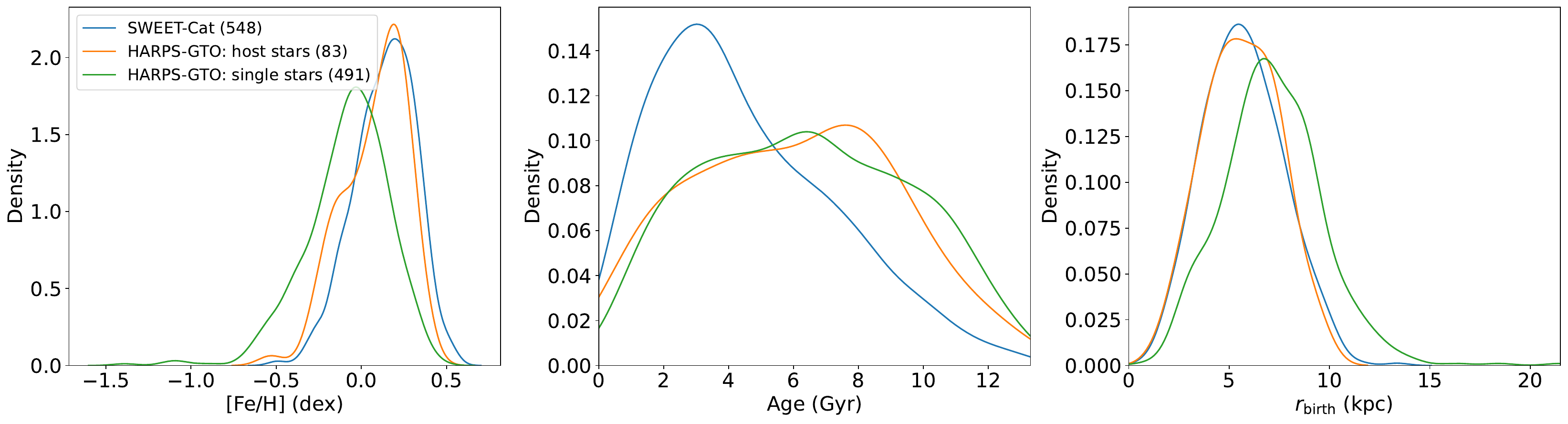}
    \caption{Distributions of [Fe/H], age, and $r_\text{birth}$ (from left to right) for the SWEET-Cat and HARPS-GTO samples. For comparison, the distributions of single stars present in the HARPS-GTO sample are plotted in all panels.} 
    \label{fig:kde}
\end{figure*}

On the leftmost panel of Fig.~\ref{fig:kde}, we show the [Fe/H] distributions for the SWEET-Cat and HARPS-GTO samples. We see that planet host stars have higher [Fe/H] than single stars. In particular, the mean [Fe/H] value is $0.14\pm0.01$ dex for the SWEET-Cat sample, $0.08\pm0.02$ for the HARPS-GTO host stars, and $-0.08\pm0.01$ for single stars. The errors represent the standard errors of the mean, which is calculated as the standard deviation divided by the square root of the sample size.
We performed the Kolmogorov-Smirnov (KS) two-sample test to statistically analyse the similarities between the three samples. For the SWEET-Cat and HARPS-GTO host stars, the KS test returns a p-value of 0.07, suggesting that these two distributions might belong to the same population.
However, the KS test results indicate that the distributions in the SWEET-Cat and HARPS-GTO host stars are statistically different from those of single stars, returning p-values of 9.54e-38 and 3.03e-07, respectively.

\begin{figure}
    \centering
    \includegraphics[scale=0.215]{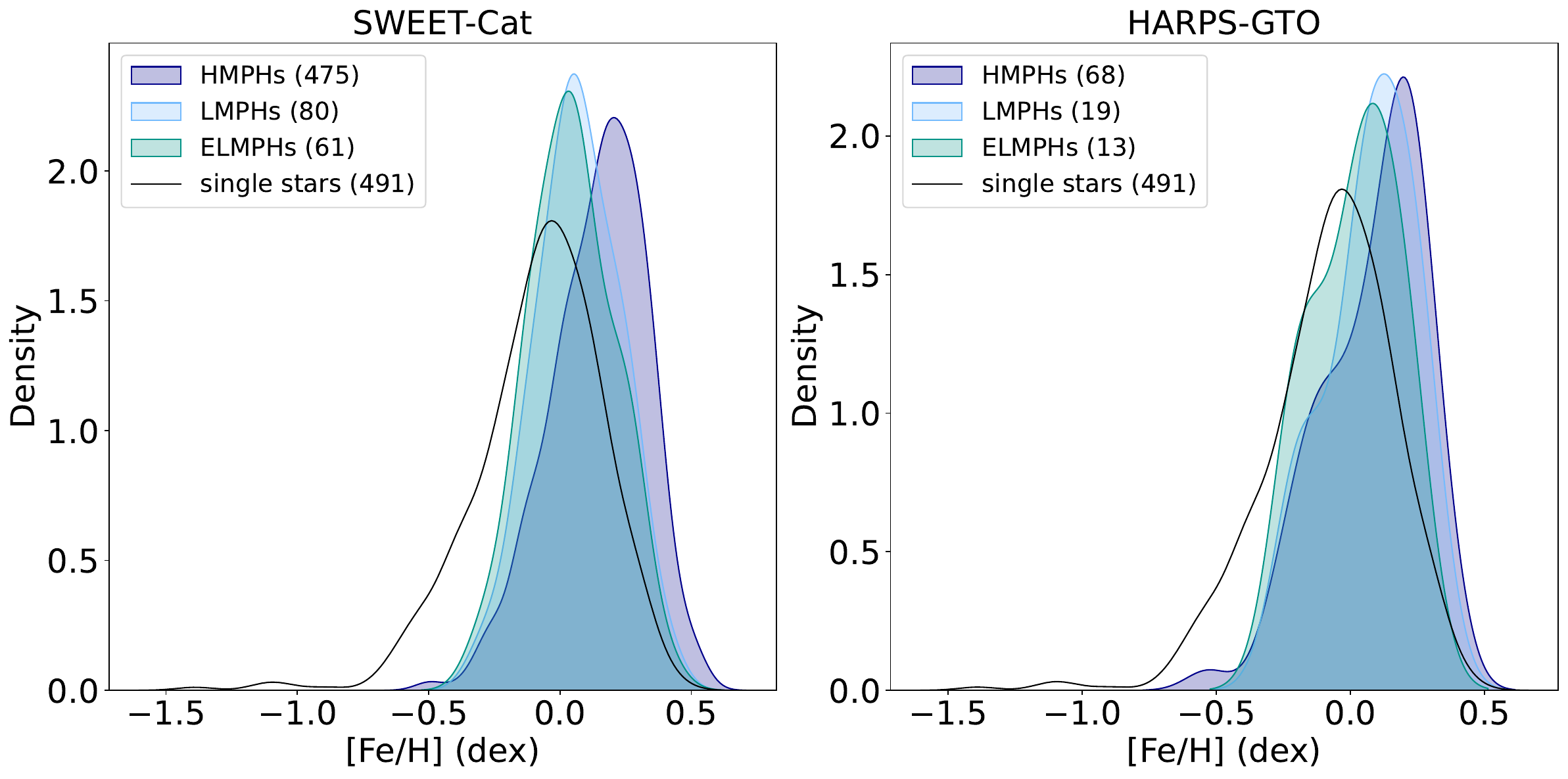}
    \caption{Distributions of [Fe/H] for HMPHs, LMPHs and ELMPHs present in the SWEET-Cat and HARPS-GTO samples. For comparison, the distributions of single stars present in the HARPS-GTO sample are plotted in both panels.}
    \label{fig:plmass-feh}
\end{figure}

The [Fe/H] distributions for HMPHs, LMPHs, and ELMPHs within the SWEET-Cat and HARPS-GTO samples are shown in Fig.~\ref{fig:plmass-feh}.
We see that HMPHs have, on average, the highest [Fe/H] of the three groups, whereas it decreases for LMPHs and ELMPHs, the latter ones being less metallic. In particular, for the SWEET-Cat sample, the mean [Fe/H] values are $0.15\pm0.01$ dex for HMPHs, $0.07\pm0.02$ dex for LMPHs, and $0.04\pm0.02$ dex for ELMPHs. For the HARPS-GTO sample, the mean [Fe/H] values are $0.09\pm0.02$ dex for HMPHs, $0.07\pm0.04$ dex for LMPHs, and $0.02\pm0.04$ dex for ELMPHs.
Table~\ref{tab:kstest} collects the p-values from the KS tests comparing the [Fe/H] distributions of HMPHs, LMPHs, ELMPHs, and single stars.

For the SWEET-Cat sample, the p-values obtained with the KS test suggest that HMPHs have statistically different [Fe/H] distributions when compared to both LMPHs and ELMPHs. On the other hand, LMPHs and ELMPHs appear to come from the same parent distribution.
In turn, HMPHs, LMPHs, and ELMPHs appear to have statistically different distributions when compared to single stars.
For the HARPS-GTO sample, HMPHs, LMPHs, and ELMPHs show no statistically significant differences between each other. When comparing single stars to HMPHs and LMPHs, the KS test returns p-values $<$ 0.05, indicating that the distributions are different, whereas single stars and ELMPHs seem to come from the same parent distribution. However, we should also note that the sample size of HARPS-GTO host stars is small.

\subsection{Age distributions} \label{age}

On the middle of Fig.~\ref{fig:kde}, we show the age distributions for the SWEET-Cat and HARPS-GTO samples. In general, we see that stars hosting planets are younger than stars without planets.
In particular, the mean age value is $4.5\pm0.1$ Gyr for the SWEET-Cat sample, $6.0\pm0.3$ Gyr for the HARPS-GTO host stars and $6.4\pm0.1$ Gyr for single stars. The KS test returns a p-value of 8.02e-05 for the SWEET-Cat and HARPS-GTO host stars and a p-value of 4.73e-17 for the SWEET-Cat and single stars, indicating that these samples do not come from the same parent population. However, for the HARPS-GTO host stars and single stars, the p-value is 0.73, indicating that these age distributions are not statistically different.

\begin{figure}
    \centering
    \includegraphics[scale=0.215]{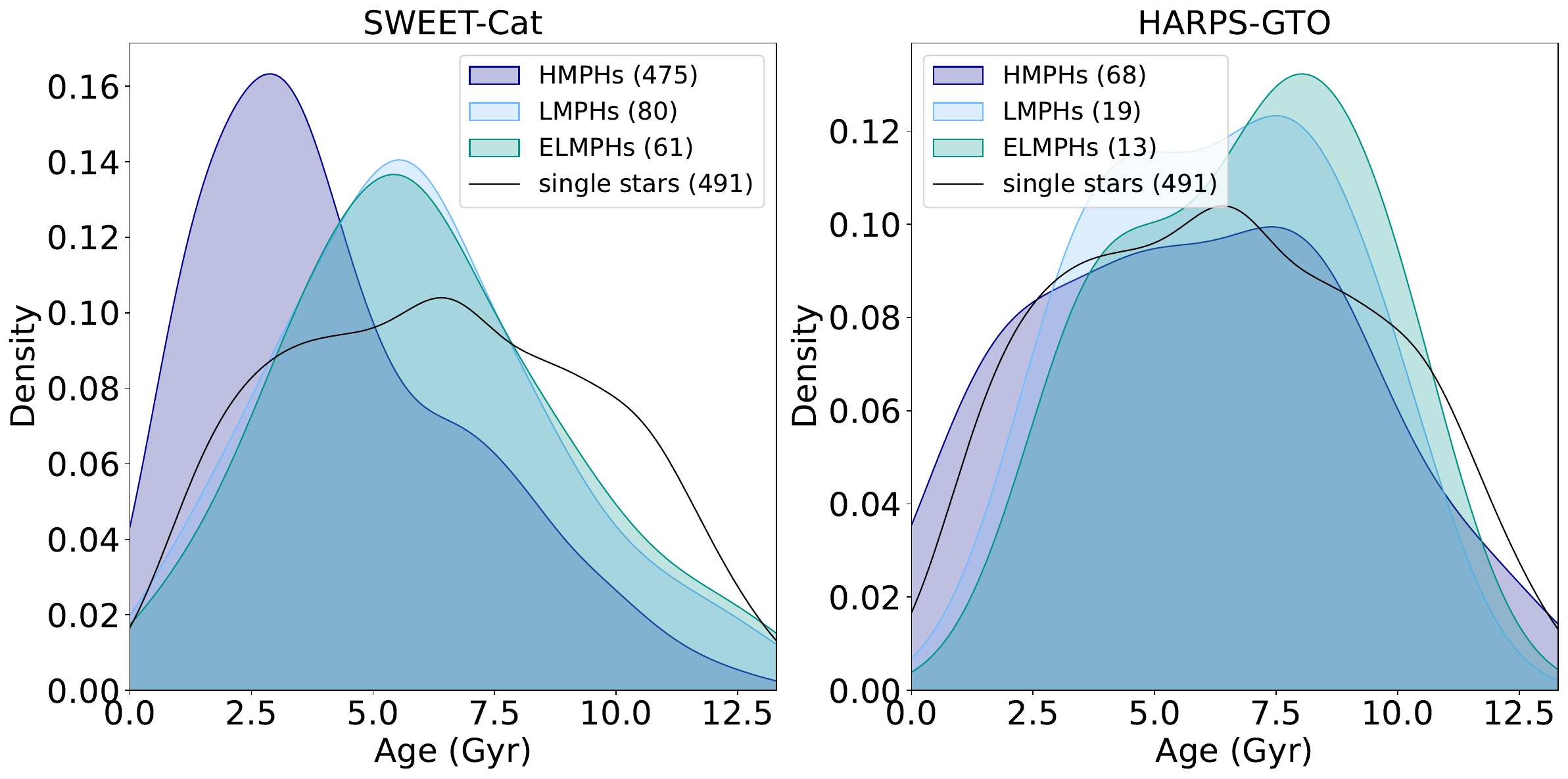}
    \caption{Age distributions for HMPHs, LMPHs and ELMPHs present in the SWEET-Cat and HARPS-GTO samples. For comparison, the distributions of single stars present in the HARPS-GTO sample are plotted in both panels.}
    \label{fig:plmass-age}
\end{figure}

The age distributions for HMPHs, LMPHs, and ELMPHs are shown in Fig.~\ref{fig:plmass-age}. Our results show that HMPHs are, on average, the youngest of the three groups. The age increases for LMPHs and ELMPHs, with the latter being older. For the SWEET-Cat sample, the mean age values are $4.3\pm0.1$ Gyr for HMPHs, $5.8\pm0.3$ Gyr for LMPHs, and $6.1\pm0.4$ Gyr for ELMPHs. Table~\ref{tab:kstest} shows the KS test p-values for the age distributions. The KS test p-values indicate that HMPHs are statistically younger than LMPHs and ELMPHs, and no differences between the age distributions of LMPHs and ELMPHs are found. On the other hand, when comparing the planetary groups with single stars, only the age distributions of HMPHs show significant differences from those of single stars.
For the HARPS-GTO sample, the mean age values are $5.9\pm0.4$ Gyr for HMPHs, $6.3\pm0.6$ Gyr for LMPHs, and $6.9\pm0.7$ Gyr for ELMPHs. We see that in this sample, ELMPHs are on average older than single stars. No significant differences are found in the age distributions between HMPHs, LMPHs, and ELMPHs, as well as when these are compared to single stars.

\subsection{Distributions of Galactic birth radii} \label{rbirth}

On the rightmost panel of Fig.~\ref{fig:kde}, we show the $r_\text{birth}$ distributions for the SWEET-Cat and HARPS-GTO samples. The results reveal that stars hosting planets have smaller $r_\text{birth}$ than stars without planets. In particular, the mean $r_\text{birth}$ value is $5.7\pm0.1$ kpc for the SWEET-Cat sample, $5.6\pm0.2$ kpc for the HARPS-GTO host stars, and $7.2\pm0.1$ kpc for single stars. When comparing SWEET-Cat stars and HARPS-GTO host stars, the KS test suggests that the $r_\text{birth}$ distributions come from the same parent distribution, providing a p-value of 0.99. However, when comparing SWEET-Cat and HARPS-GTO host stars with single stars, it returns p-values of 1.01e-17 and 1.88e-05, respectively, confirming that the distributions are statistically different.

\begin{figure}
    \centering
    \includegraphics[scale=0.215]{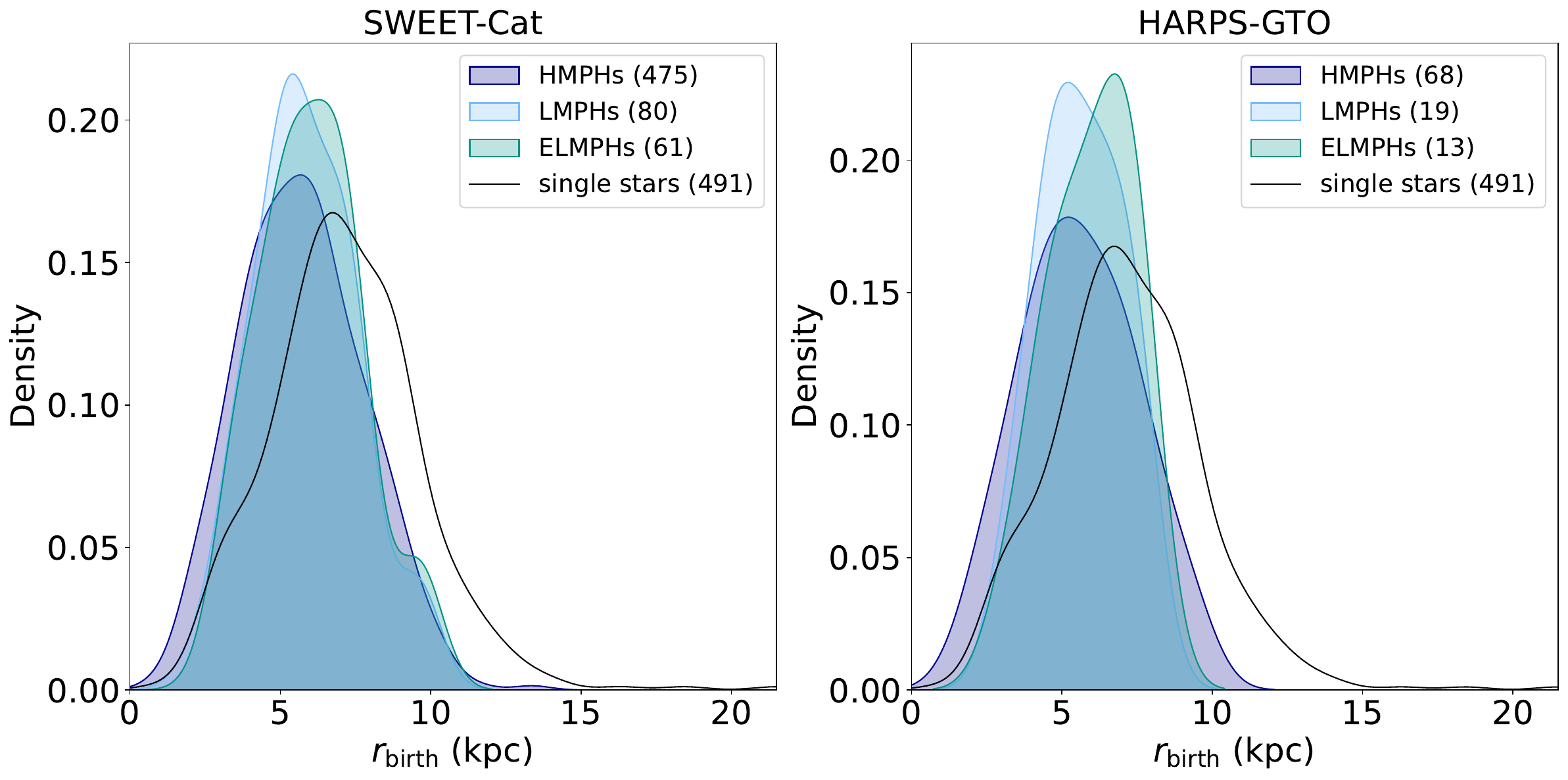}
    \caption{Distributions of $r_\text{birth}$ for HMPHs, LMPHs and ELMPHs present in the SWEET-Cat and HARPS-GTO samples. For comparison, the distributions of single stars present in the HARPS-GTO sample are plotted in both panels.}
    \label{fig:plmass-rbirth}
\end{figure}

The $r_\text{birth}$ distributions for HMPHs, LMPHs, and ELMPHs are shown in Fig.~\ref{fig:plmass-rbirth}. The results reveal that HMPHs have, on average, smaller $r_\text{birth}$ than LMPHs and ELMPHs, the latter having the smallest $r_\text{birth}$ among the three groups. In particular, the mean $r_\text{birth}$ values in the SWEET-Cat sample are $5.7\pm0.1$ kpc for HMPHs, $5.9\pm0.2$ kpc for LMPHs, and $6.1\pm0.2$ kpc for ELMPHs. In the HARPS-GTO sample, the mean $r_\text{birth}$ values are $5.6\pm0.2$ kpc for HMPHs, $5.6\pm0.3$ kpc for LMPHs, and $6.0\pm0.4$ kpc for ELMPHs. Table~\ref{tab:kstest} shows the KS test p-values for the $r_\text{birth}$ distributions.
When comparing the $r_\text{birth}$ distributions of HMPHs, LMPHs, and ELMPHs, we find no statistically significant differences in both the SWEET-Cat and HARPS-GTO samples. However, in the SWEET-Cat sample, the difference between the planetary groups and single stars is always significant. In the HARPS-GTO sample, only ELMPHs and single stars seem to come from the same parent distribution.

\begin{table}
\caption{The results of the KS tests comparing the [Fe/H], age, and $r_\text{birth}$ distributions for the HMPHs, LMPHs, ELMPHs, and single stars.}
\label{tab:kstest}
\resizebox{\columnwidth}{!}{%
\begin{tabular}{llll}
\hline
                               & \multicolumn{3}{c}{KS test p-value}                                                           \\ \hline
Samples                        & \multicolumn{1}{c}{{[}Fe/H{]}} & \multicolumn{1}{c}{Age} & \multicolumn{1}{c}{$r_\text{birth}$} \\ \hline
\textit{SWEET-Cat}             &                                &                         &                            \\
HMPHs vs. LMPHs                 & \textbf{4.00e-04}                        & \textbf{4.81e-06}                & 0.35                       \\
HMPHs vs ELMPHs                & \textbf{3.81e-05}                       & \textbf{1.40e-05}                & 0.27                       \\
LMPHs vs. ELMPHs                & 0.95                           & 0.99                    & 0.99                       \\
HMPHs vs. single stars & \textbf{3.64e-39}                       & \textbf{1.95e-19}                & \textbf{6.64e-17}                   \\
LMPHs vs. single stars & \textbf{7.53e-06}                        & 0.13                    & \textbf{1.16e-05}                   \\
ELMPHs vs. single stars & \textbf{3.00e-3}                         & 0.42                    & \textbf{3.00e-4}                     \\ \hline
\textit{HARPS-GTO}             &                                &                         &                            \\
HMPHs vs. LMPHs                 & 0.87                           & 0.63                    & 0.79                       \\
HMPHs vs. ELMPHs                & 0.27                           & 0.50                     & 0.60                       \\
LMPHs vs. ELMPHs                & 0.87                           & 0.99                    & 0.98                       \\
HMPHs vs. single stars & \textbf{1.27e-07}                       & 0.63                    & \textbf{2.20e-05}                   \\
LMPHs vs. single stars & \textbf{4.00e-03}                         & 0.76                    & \textbf{0.01}                       \\
ELMPHs vs. single stars & 0.16                           & 0.48                    & 0.05                       \\ \hline
\end{tabular}%
}
\end{table}

\subsection{Formation efficiency of planetary systems} \label{freq}

In this subsection, we analyse the frequency of planetary systems as a function of [Fe/H], age, and $r_\text{birth}$.
We define the relative frequency of planetary systems, $F_\text{p}$, simply as
\begin{equation}
    F_\text{p} = \frac{\text{Total number of planetary systems}}{\text{Total number of stars}},
\end{equation}
where the total number of stars is the sum of the number of stars with and without planets.
We computed the $F_\text{p}$ for HMPHs and LMPHs for both the SWEET-Cat and HARPS-GTO samples, and we interpreted it as the formation efficiency of high- and low-mass planets, respectively. Although the HARPS-GTO is a volume-limited sample, we note that this is not the mathematically correct way to compute the frequency of planetary systems, as corrections for the incompleteness of the samples should be taken into account \citep[e.g.,][]{Faria2016,Santerne2016}.
Therefore, these numbers do not represent the true planetary frequency. Nevertheless, the observed trends should reflect the true dependencies of the planet formation efficiency on [Fe/H], age, and $r_\text{birth}$.

\begin{figure}
    \centering
    \includegraphics[scale=0.205]{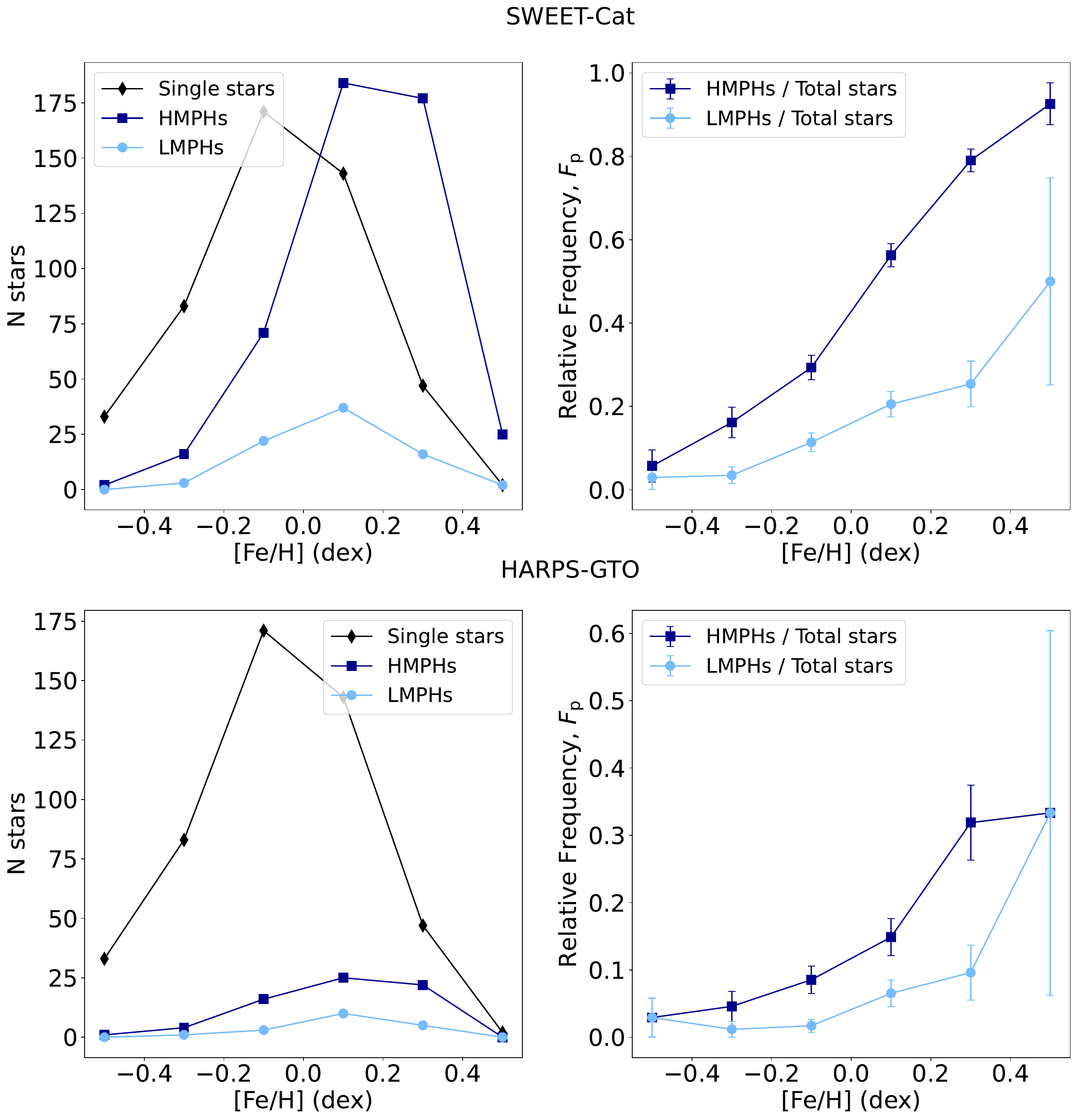} 
    \caption{Left panel: Number of HMPHs, LMPHs, and single stars as a function of [Fe/H]. Right panel: Relative frequency of HMPHs and LMPHs as a function of [Fe/H].}
    \label{fig:frequency_feh}
\end{figure}

In Fig.~\ref{fig:frequency_feh} is shown the number of HMPHs, LMPHs, and single stars (left panel) and the relative frequency of HMPHs and LMPHs (right panel) as a function of [Fe/H] for both the SWEET-Cat and HARPS-GTO samples. The upper and lower uncertainties were computed using a binomial distribution and corresponded to a confidence interval level of 68\%. It is observed that the formation efficiency of planets increases with [Fe/H]. For the SWEET-Cat sample, the formation efficiency of high-mass planets is higher and increases steeply with [Fe/H] compared to the formation efficiency of low-mass planets.
For the HARPS-GTO sample, we see the same relation. In both samples, for low-mass planets, we observe a steep increase around [Fe/H] > +0.3 dex, which is likely due to the small sample size of LMPHs.

\begin{figure}
    \centering
    \includegraphics[scale=0.205]{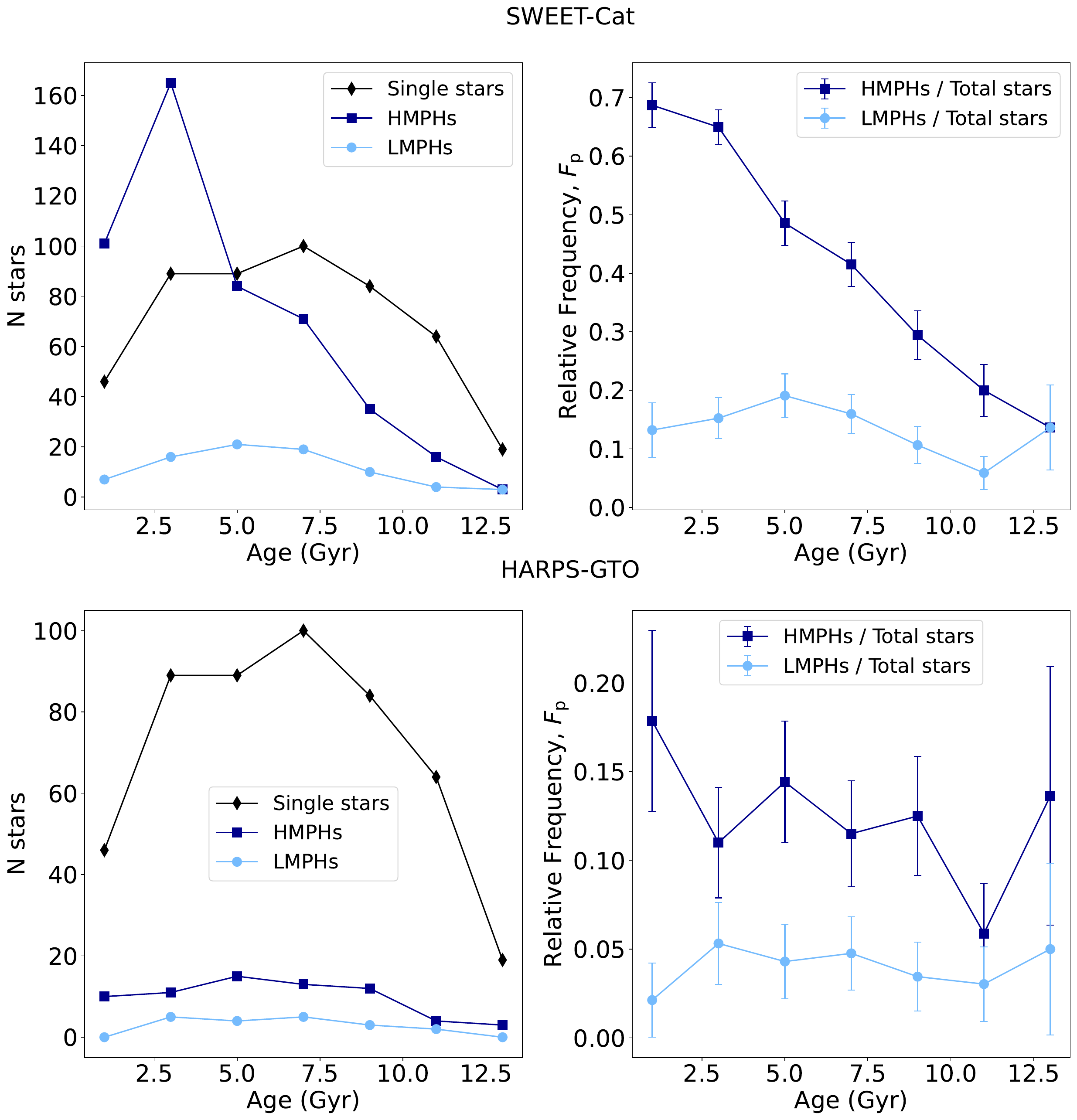} 
    \caption{Left panel: Number of HMPHs, LMPHs, and single stars as a function of age. Right panel: Relative frequency of HMPHs and LMPHs as a function of age.}
    \label{fig:frequency_age}
\end{figure}

Figure~\ref{fig:frequency_age} shows the same as Fig.~\ref{fig:frequency_feh}, but for stellar age.
For the SWEET-Cat sample, the results show that the formation efficiency of high-mass planets decreases steeply with age. On the other hand, the formation efficiency of low-mass planets is lower than that of high-mass planetary systems, for which we observe a slight increase towards $\sim$ 5 Gyr and then a decrease until $\sim$ 11 Gyr. For ages > 11 Gyr, the formation efficiency of these planets shows an increase, but this is due to the small sample size of LMPHs and single stars, as indicated by the large uncertainties.
For the HARPS-GTO sample, we see that the formation efficiency with stellar age increases more gradually for high-mass planets and is almost constant for low-mass planets.

\begin{figure}
    \centering
    \includegraphics[scale=0.205]{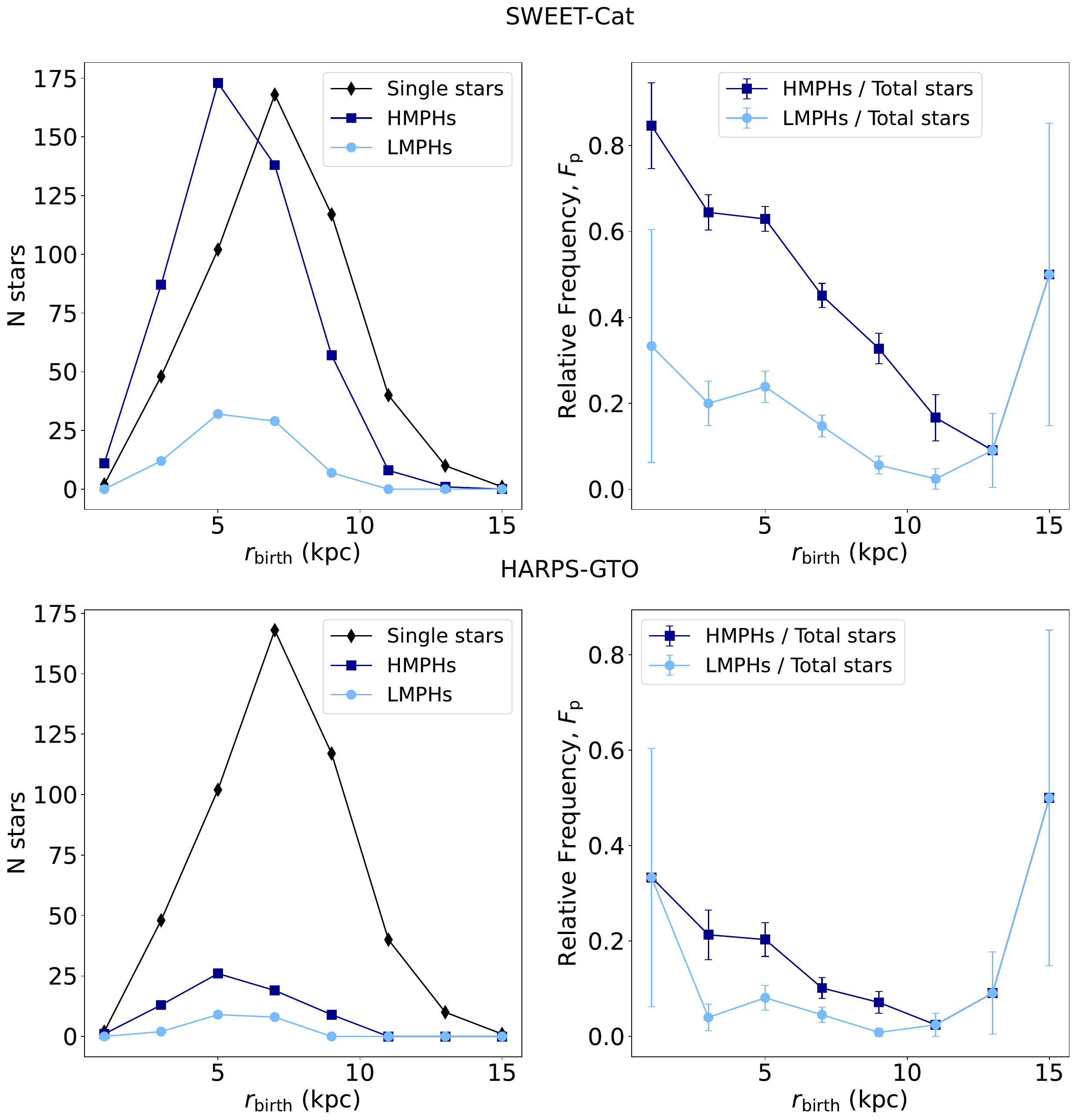}
    \caption{Left panel: Number of HMPHs, LMPHs, and single stars as a function of $r_\text{birth}$. Right panel: Relative frequency of HMPHs and LMPHs as a function of $r_\text{birth}$.}
    \label{fig:frequency_rbirth}
\end{figure}

Figure~\ref{fig:frequency_rbirth} shows the same as Fig.~\ref{fig:frequency_feh}, but for $r_\text{birth}$. In general, we see that the formation efficiency of planets decreases with galactocentric distance. 
For the SWEET-Cat sample, the formation efficiency of high-mass planets decreases steeply with $r_\text{birth}$ and decreases more gradually for low-mass planets. For the HARPS-GTO sample, we observe the same trend for high-mass planets, although it is more gradual. However, for low-mass planets, the formation efficiency appears nearly constant with $r_\text{birth}$, with the exception of $r_\text{birth} < 2$ kpc, where the sample size of LMPHs is small.
For $r_\text{birth} > 11$ kpc, we see an increase in the formation efficiency of planets for both HMPHs and LMPHs in both the SWEET-Cat and HARPS-GTO samples, as indicated by the large error bars.

\subsection{Time evolution of the Galactic birth radii} \label{evo}

\begin{figure*}
     \centering
     \includegraphics[scale=0.31]{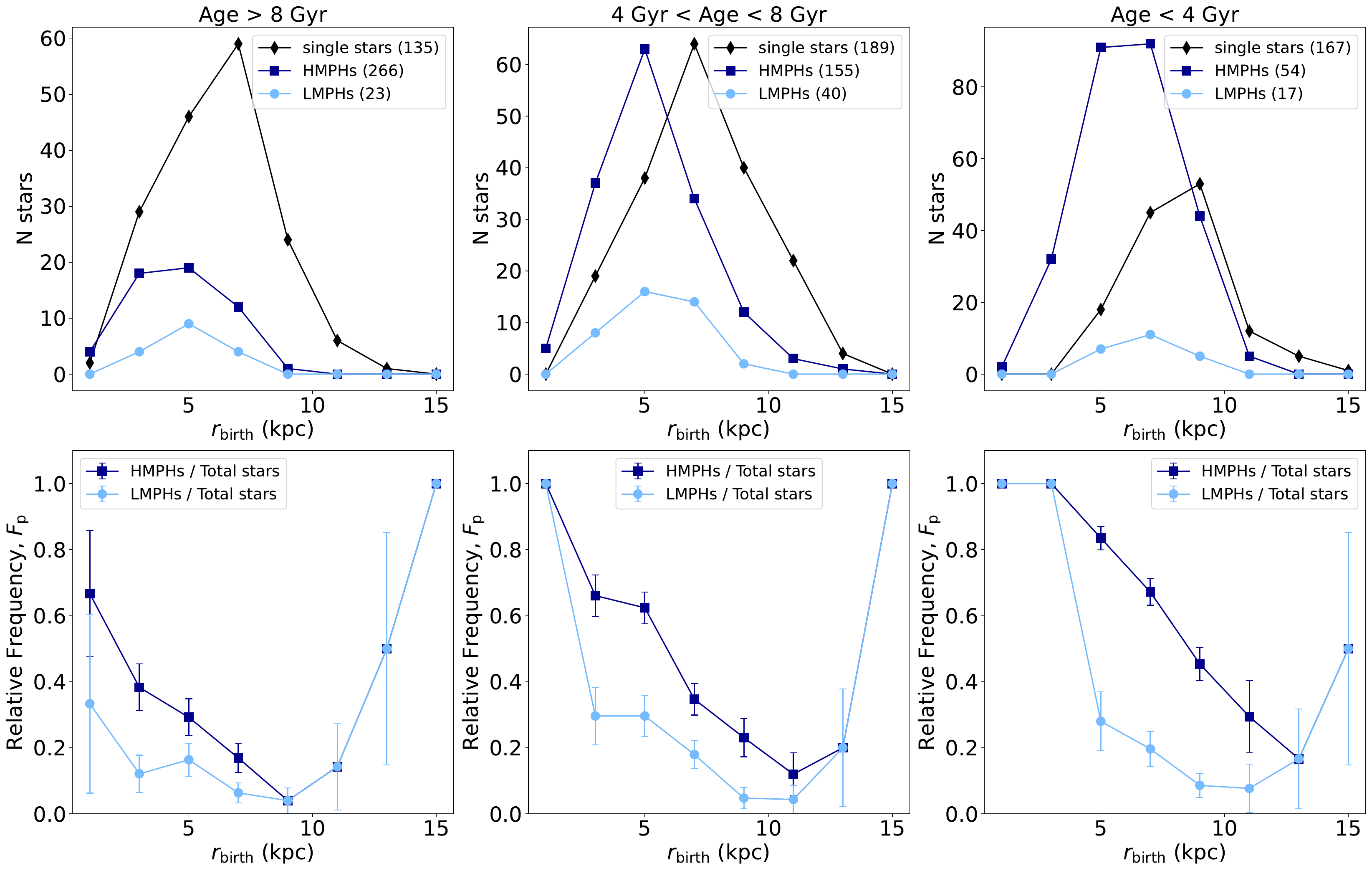} 
     \caption{Top panel: Number of HMPHs, LMPHs, and single stars with ages above 8 Gyr (left), between 4 and 8 Gyr (middle), and below 4 Gyr (right), as a function of $r_\text{birth}$. Bottom panel: Relative frequency of HMPHs and LMPHs with ages above 8 Gyr (left), between 4 and 8 Gyr (middle), and below 4 Gyr (right), as a function of $r_\text{birth}$.
     }
     \label{fig:SC_age_birth}
 \end{figure*}

To better understand these results and how the $r_\text{birth}$ of stars hosting planets changes over time, we show in Fig.~\ref{fig:SC_age_birth} the time evolution of the $r_\text{birth}$, focusing only on the SWEET-Cat sample, since the number of planets detected in the HARPS-GTO sample is small.

In the top leftmost panel of Fig.~\ref{fig:SC_age_birth}, we observe that at early times (age > 8 Gyr), the number of planet host stars is lower compared to stars without detected planets. The mean $r_\text{birth}$ values for this age range are $4.7\pm0.3$ kpc for HMPHs, $5.1\pm0.3$ kpc for LMPHs, and $6.2\pm0.2$ kpc for single stars. In the bottom leftmost panel, the formation efficiency of high-mass planets shows a steep decrease with galactocentric distance until $\sim 9$ kpc, whereas the formation efficiency of low-mass planets decreases more gradually.

In the top middle panel of Fig.~\ref{fig:SC_age_birth}, we see that at intermediate ages (4 Gyr < age < 8 Gyr), the number of stars with planets increases significantly compared to the earlier times. The mean $r_\text{birth}$ values for this age domain are $5.3\pm0.2$ kpc for HMPHs, $5.6\pm0.3$ kpc for LMPHs, and $7.3\pm0.2$ kpc for single stars. In the bottom middle panel, we observe that the formation efficiency of high- and low-mass planets increases at any radius compared to ages > 8 Gyr. 

In the top rightmost panel of Fig.~\ref{fig:SC_age_birth}, at later times (age < 4 Gyr), the number of HMPHs increases significantly compared to other ages encompassing larger galactocentric distances, reaching a mean $r_\text{birth}$ value of $6.2\pm0.1$ kpc. The number of LMPHs shows a decrease, encompassing a mean $r_\text{birth}$ of $7.0\pm0.3$ kpc. On the other hand, the number of single stars decreases significantly and encompasses a mean $r_\text{birth}$ of $8.1\pm0.2$ kpc. Furthermore, we see in the bottom rightmost panel that the formation efficiency of high-mass planets shows an increase at any radius compared to other ages, while that of low-mass planets decreases at any radius. 

In the bottom panels of Fig.~\ref{fig:SC_age_birth}, we notice a sudden increase in the formation efficiency of both high- and low-mass planets at larger galactocentric radii, which is explained by the low number of stars with and without planets at these radii.

\section{Discussion and conclusion}          \label{sec:conclusion}

In this work, we inferred the Galactic birth radii of exoplanets in order to understand the Galactic aspects of exoplanet formation and evolution. We used a sample of stars hosting planets from the SWEET-Cat catalogue and a sample of stars with and without planets from the HARPS-GTO database. We employed the same method as \cite{Minchev} to infer the Galactic birth radii using only precise [Fe/H] and age estimations.

Our results show that stars hosting planets statistically have higher [Fe/H], are younger, and have smaller $r_\text{birth}$ than stars without detected planets.

In particular, from the SWEET-Cat sample, our results suggest that stars hosting high-mass planets have different [Fe/H] and age distributions compared to stars hosting at least one low-mass planet and stars hosting exclusively low-mass planets. The results indicate that high-mass planets preferentially form and are detected around higher [Fe/H] and younger stars compared to low-mass planets (Sect.~\ref{feh} and Sect.~\ref{age}).
Additionally, the results show that systems with both high- and low-mass planets have, on average, higher [Fe/H] and are younger compared to systems with only low-mass planets. Although the KS test results did not find that these distributions are statistically different between these groups in both samples, the formation of both low-mass and high-mass planets in the same system may only occur at later times when the ISM is sufficiently enriched with heavier elements.

We show in Section \ref{freq} that the formation efficiency increases steeply with [Fe/H] for high-mass and gradually for low-mass planets. It is observed that the formation efficiency decreases with age for high-mass planets. On the other hand, the formation efficiency of low-mass planets is less dependent on age, being slightly higher at the ages between 4 and 8 Gyr, which is shown as well in Sect.~\ref{evo}.
However, it is important to note that the detection of low-mass planets around young stars is very challenging due to the high stellar activity present in these stars. As such, some low-mass planets may not have been detected around these stars yet.

Regarding the HARPS-GTO sample results, the KS test p-values indicate that the HARPS-GTO host stars are statistically older than the SWEET-Cat stars. However, this difference does not occur when comparing the [Fe/H] distributions.
This is probably because of detection bias. This may be explained by the fact that the planets constituting the HARPS-GTO sample were primarily detected with the radial velocity method, and thus more challenging is detecting planets around younger stars. Furthermore, the sample size of HARPS-GTO host stars is too small to draw a firm conclusion.

In general, these findings are consistent with the results present in the literature. Recently, \cite{Nielsen2023} found that stars with increasing metallicity form planets with increasing masses for both giant planets and rocky planets.
\cite{Swastik2022, Swastik2023,Swastik2024} found that high-mass planetary systems are statistically younger than low-mass planetary systems, which is a likely consequence of the Galactic chemical evolution, in which at later times the ISM was enriched with Fe-peak elements, enabling the formation of high-mass planets. Regarding the occurrence of low-mass planets, \cite{Lu2020} found that the occurrence rate of close-in small planets showed a slight dependence on stellar metallicity, and \cite{Sousa2019} found that the mass of low-mass planets increases with metallicity and orbital period.

Regarding the Galactic birth radii distributions (Sect.~\ref{rbirth}), we found that stars hosting planets have a statistically smaller $r_\text{birth}$ compared to their non-host counterparts.
No significant statistical difference is found between the $r_\text{birth}$ distributions of the planetary host star groups in both the SWEET-Cat and HARPS-GTO samples. However, stars hosting high-mass planets show, on average, smaller $r_\text{birth}$ than stars hosting low-mass planets and stars hosting exclusively low-mass planets. This is also validated by the results of the formation efficiency of high-mass planets, which is strongly dependent on $r_\text{birth}$ compared to low-mass planets.

These findings are consistent with the \cite{Baba2023} Galactic chemical evolution model, which predicted a higher occurrence rate of gas giant planets with time and towards the inner Galactic regions. More recently, \cite{Boettner2024} applied their model to a simulated Milky Way analogue and found that the planet demographics in the central, metal-rich regions of the Milky Way analogue differ strongly from the planet populations in the more distant, metal-poor regions, with the occurrence rate of giant planets significantly higher in the thin disc than in the thick disc. On the other hand, they expect low-mass planets to be abundant throughout the Galaxy since their formation processes are less dependent on stellar metallicity.
Earlier, \cite{Haywood2009} suggested that metal-rich stars hosting giant planets originate from the inner disk.

To conclude, in Section~\ref{evo}, we examined the time evolution of the Galactic birth radii. Our results suggest that the formation efficiency of high-mass planets increases with time and encompasses a larger galactocentric distance over time. 
These results are in agreement with (i) the observed negative ISM metallicity gradient, where at small galactocentric distances, the metallicity is higher, favouring the formation of these planets, as predicted by core accretion models (ii) the ISM enrichment and flattening with time at any radius, which favours the formation of these planets around younger and metal-richer stars, encompassing larger galactocentric radii.
The formation efficiency of low-mass planets shows a slight increase between the ages of 4 and 8 Gyr and also encompasses larger $r_\text{birth}$ over time.
Stars without detected planets show to form at larger galactocentric distances with time, as the ISM at larger Galactic distances becomes sufficiently enriched with time to form stars but not planets.

Our analysis serves as an initial step toward understanding the galactic aspects of planet formation. With the upcoming PLATO mission \citep{PLATO}, we anticipate a significant increase in both the sample size and, more importantly, the accuracy of age determinations—currently one of the key limiting factors in our study. These advancements will pave the way for a deeper understanding of the connection between stellar populations and planetary systems across the Galaxy.


\section*{Acknowledgments}

This work was supported by the grant CIAAUP-11/2023-BI-M under the project EXO-Terra (2022.06962.PTDC) funded by Funda\c{c}\~ao para a Ci\^encia e Tecnologia. V.A. acknowledges funding support by Funda\c{c}\~ao para a Ci\^encia e Tecnologia through national funds and by FEDER through COMPETE2020 - Programa Operacional Competitividade e Internacionalização by these grants: UIDB/04434/2020; UIDP/04434/2020; 2022.06962.PTDC and by the European Union (ERC, FIERCE, 101052347). D.B. acknowledges funding support by the Italian Ministerial Grant PRIN 2022, ``Radiative opacities for astrophysical applications'', no. 2022NEXMP8, CUP C53D23001220006. This research has made use of the NASA Exoplanet Archive operated by the California Institute of Technology under contract with the National Aeronautics and Space Administration under the Exoplanet Exploration Program.





\subsection*{Conflicts of interest}

The authors declare no conflicts of interest.

\nocite{*}
\bibliography{Wiley-ASNA}%

\end{document}